\documentclass[prl,twocolumn,showpacs,amsmath,amssymb]{revtex4}

\usepackage{graphicx}%Include figure files
\usepackage{dcolumn}%Align table columns on decimal point
\usepackage{bm}% bold math
\usepackage{epsfig}

\begin{document}

\title{Spin and Charge Shot Noise in Mesoscopic Spin Hall Systems}

\author{Ralitsa L. Dragomirova}
\author{Liviu P. Z\^ arbo}
\altaffiliation[Present address: ]{Department of Physics, Texas A\&M University, College Station, TX 77843-4242, USA}
\author{Branislav K. Nikoli\' c}
\affiliation{Department of Physics and Astronomy, University
of Delaware, Newark, DE 19716-2570, USA}

\begin{abstract} Injection of unpolarized charge current through the longitudinal leads of a
four-terminal two-dimensional electron gas with the Rashba spin-orbit (SO) coupling and/or SO
scattering off extrinsic impurities is responsible not only for the pure spin Hall current in the
transverse leads, but also for random time-dependent current fluctuations. We employ the scattering
approach to current-current correlations in multiterminal nanoscale conductors to analyze the shot noise of
transverse pure spin Hall and zero charge current, or transverse spin current and non-zero charge
Hall current, driven by unpolarized or spin-polarized longitudinal current, respectively. Since any
spin-flip acts as an  additional source of noise, we argue that these shot noises offer a unique tool
to differentiate between intrinsic and extrinsic SO mechanisms underlying the spin Hall effect
in paramagnetic devices.
\end{abstract}

\maketitle

{\em Introduction}---The recently discovered spin Hall effect (SHE) in paramagnetic
semiconductor~\cite{Kato2004b,Awschalom2007} and metallic~\cite{Valenzuela2006} systems holds great promise to
revolutionize~\cite{Awschalom2007} electrical generation, control, and detection of nonequilibrium spin populations in the
envisaged spintronic devices. The SHE actually denotes a {\em collection} of phenomena manifesting as transverse (with respect
to injected conventional unpolarized charge current) separation of spin-$\uparrow$ and spin-$\downarrow$ states, which then comprise
either a pure spin current or accumulate at the lateral sample boundaries. Its Onsager reciprocal effect---the inverse
spin Hall effect~\cite{Hirsch1999} where longitudinal pure spin current generates transverse charge current or voltage between
the lateral boundaries---offers one of the most efficient schemes to detect elusive {\em pure} (i.e.,
not accompanied by any net charge flux) spin currents by converting them into electrical quantities~\cite{Valenzuela2006}.

While the direct and inverse SHE do not require magnetic field, they essentially rely on
SO coupling effects in solids. In addition, their magnitude typically depends on the type of
microscopic SO interaction, impurities, charge density, geometry, and dimensionality. Such a variety of
SHE manifestations poses immense challenge for attempts at a unified theoretical description of spin transport in
the presence of relativistic effects, which has not been resolved by early hopes~\cite{Sinova2004,Guo2008} that
auxiliary spin current ${\hat j}^z_y$ and spin conductivity $\sigma_{sH}=\langle {\hat j}^z_y
\rangle/E_x$ (as the linear response to longitudinal electric field $E_x$) of infinite homogeneous systems could be
elevated to universally applicable and experimentally relevant quantities.

Thus, the key task emerging for theoretical analysis is to provide guidance for increasing  and controlling the spin accumulation in confined geometries~\cite{Nikoli'c2006a,Zyzin2008} (observed SHE in semiconductors is presently rather small~\cite{Kato2004b,Awschalom2007}) or outflowing spin currents~\cite{Nikoli'c2006a,Nikoli'c2005e} driven by them . In this respect, understanding of the intrinsic~\cite{Sinova2004,Guo2008} (due to SO-governed spin-split band structure) or extrinsic~\cite{Hirsch1999} (due to SO-dependent scattering off impurities) origin of the SHE has been one of the central topics in interpreting experiments~\cite{Guo2008} and development of  SHE-based spintronic devices~\cite{Awschalom2007}. The intrinsic SO couplings are predicted to yield large SHE response~\cite{Guo2008}, which, moreover, can be controlled electrically by gate electrodes covering low-dimensional devices~\cite{Nikoli'c2005e}. The extrinsic ones are fixed and the corresponding much smaller SHE is hardly controllable (except through charge density and mobility~\cite{Awschalom2007}).

However, measurements of usual  quantities of transverse spin and charge transport typically are not able to resolve the intrinsic vs.
extrinsic dilemma. This long standing issue is well-known from the studies of the anomalous Hall effect (AHE)~\cite{Sinitsyn2008} in ferromagnetic materials (SHE can be viewed as the zero magnetization limit of AHE) where standard analysis of the AHE experimental data---fitting of the Hall resistivity vs. longitudinal resistivity by a power law~\cite{Kotzler2005}---is insufficient to clearly differentiate underlying SO mechanisms. Here lessons from mesoscopic quantum physics might be enlightening: much more information, when contrasted to usual time-averaged
conductances and conductivities, about transport of non-interacting or interacting quasiparticles is contained in time-dependent
nonequilibrium current (or voltage) fluctuations~\cite{Blanter2000}. Unlike equilibrium (thermally driven) noise, such {\em shot noise} persists down to zero temperature and it is fundamentally connected to the discrete nature of the electron charge~\cite{Blanter2000}. Furthermore, a handful of recent theoretical~\cite{Mishchenko2003,Dragomirova2007} and experimental~\cite{Guerrero2006}
studies have suggested that shot noise in systems with spin-dependent interactions provides a sensitive probe to differentiate between magnetic impurities, spin-flip scattering, and continuous spin precession effects in transport. This is due to the fact that any spin flip introduces additional source of current fluctuation when spin degeneracy is lifted and electrons from spin-$\uparrow$ subsystem are converted into spin-$\downarrow$ subsystem.

\begin{figure}
\centerline{\psfig{file=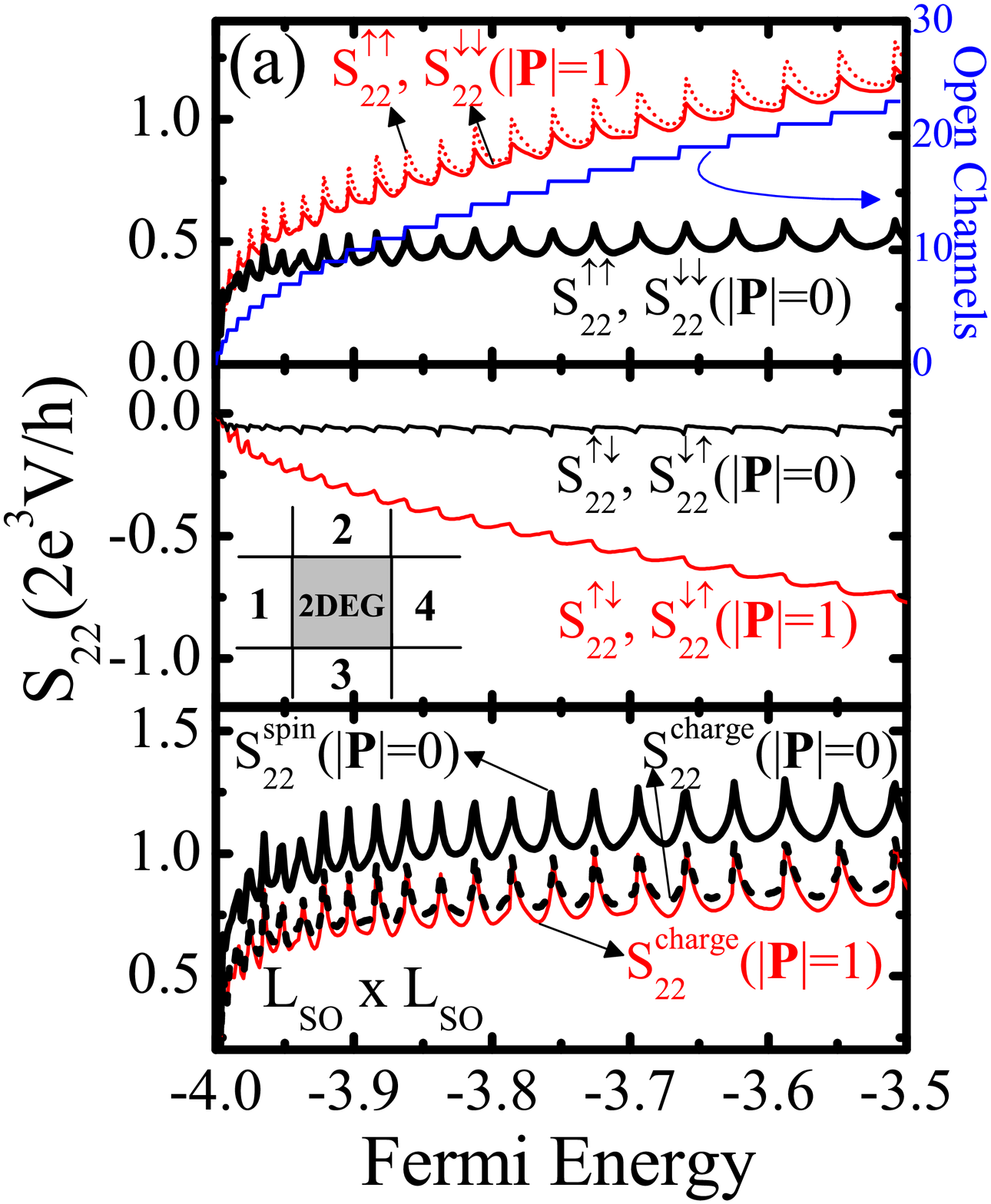,scale=0.21,angle=0} \hspace{-0.1in} \psfig{file=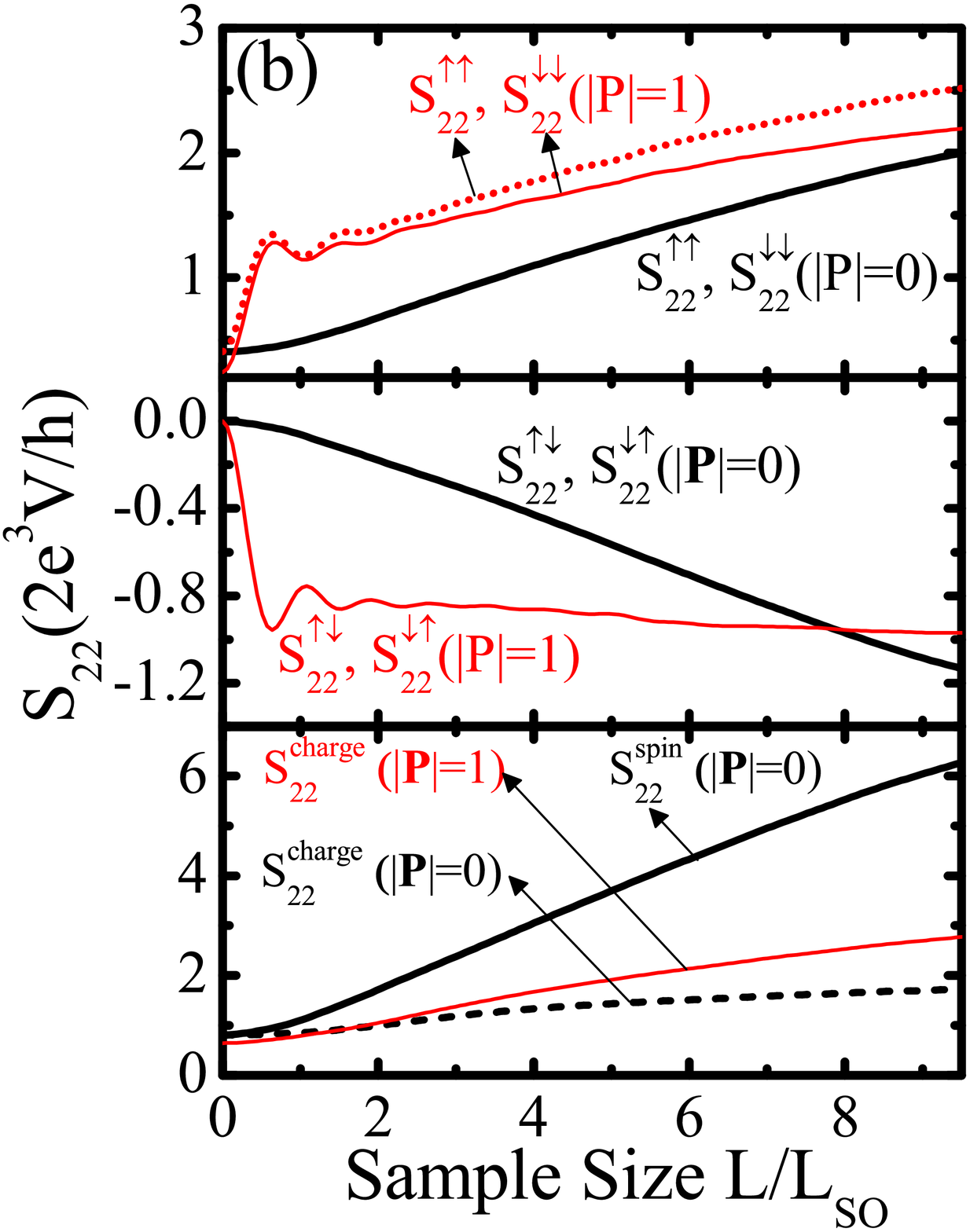,scale=0.21,angle=0}}
\caption{(Color online) Ballistic spin-resolved shot noise  in the transverse electrode 2 [see inset in panel (a)], as well as the total shot 
noise of pure spin Hall current (driven by unpolarized $|{\bf P}|=0$ injected charge current $I_1$) or charge Hall current [driven by spin-polarized ${\bf P}=(0,0,1)$ injection of $I_1$], in {\em clean} 2DEGs with the Rashba SO coupling as a function of: (a) Fermi energy $E_F$; or (b) Rashba SO coupling $\alpha$ measured through the spin precession length $L_{\rm SO}=\pi \hbar^2/2m^* \alpha$. In panel (a) the 2DEG sample is of the size $L_{\rm SO} \times L_{\rm SO}$, while in panel (b) the sample size is $300 \, {\rm nm} \times 300 \, {\rm nm}$ and $E_F$ is fixed to open 23 channels for electron injection from lead 1.}\label{fig:ballistic}
\end{figure}

\begin{figure*}
\centerline{\psfig{file=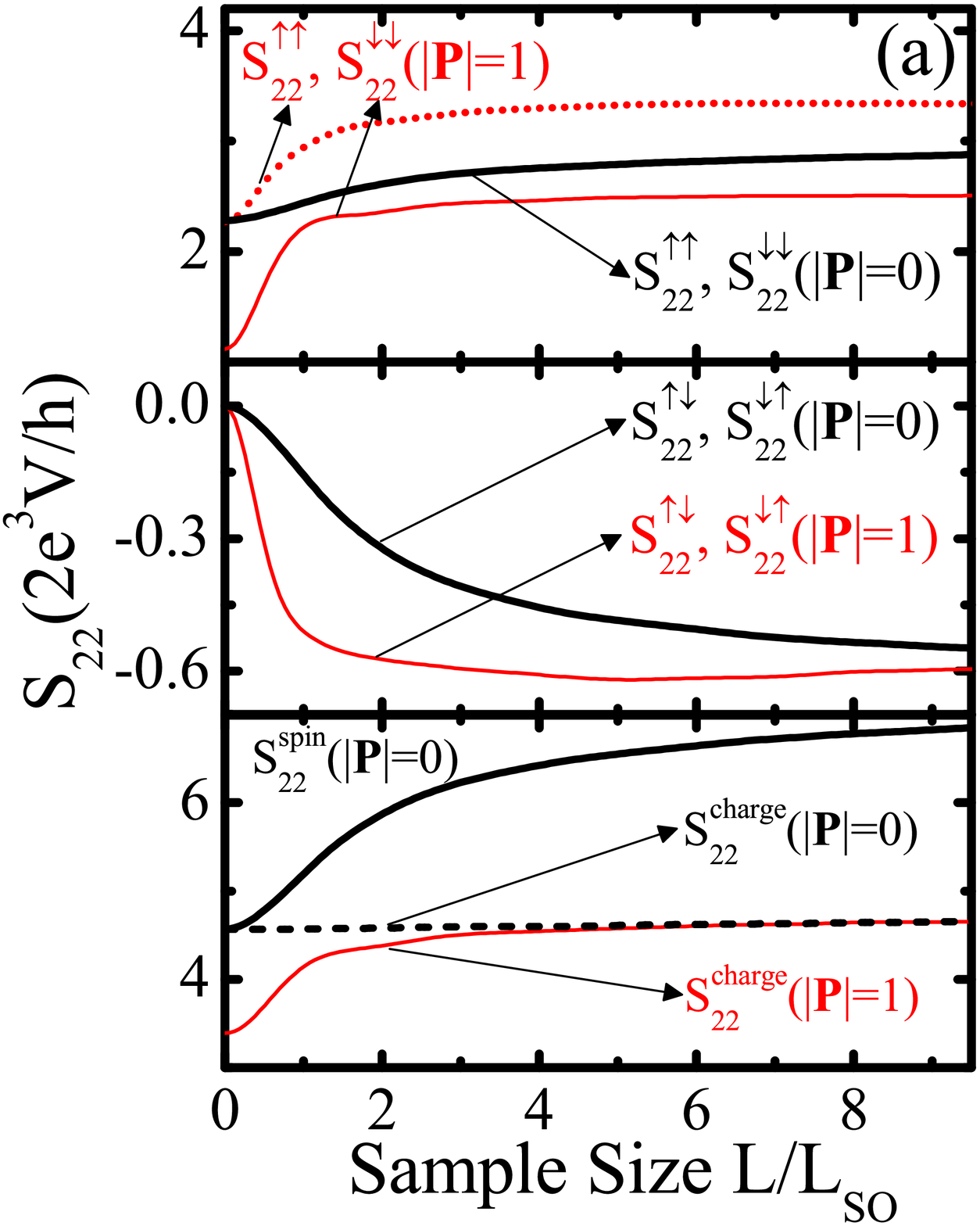,scale=0.22,angle=0} \hspace{0.2in} \psfig{file=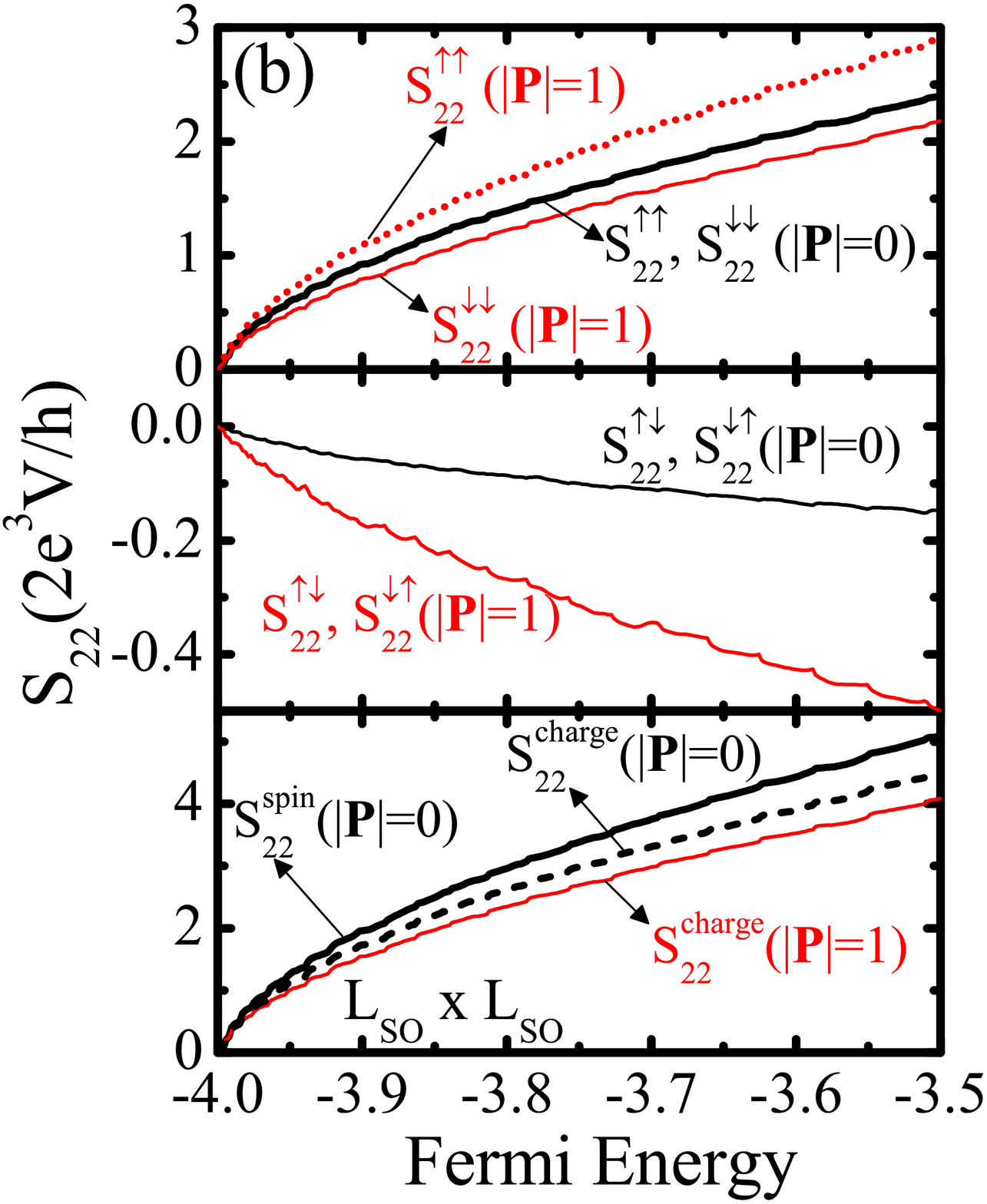,scale=0.22,angle=0} \hspace{0.2in} \psfig{file=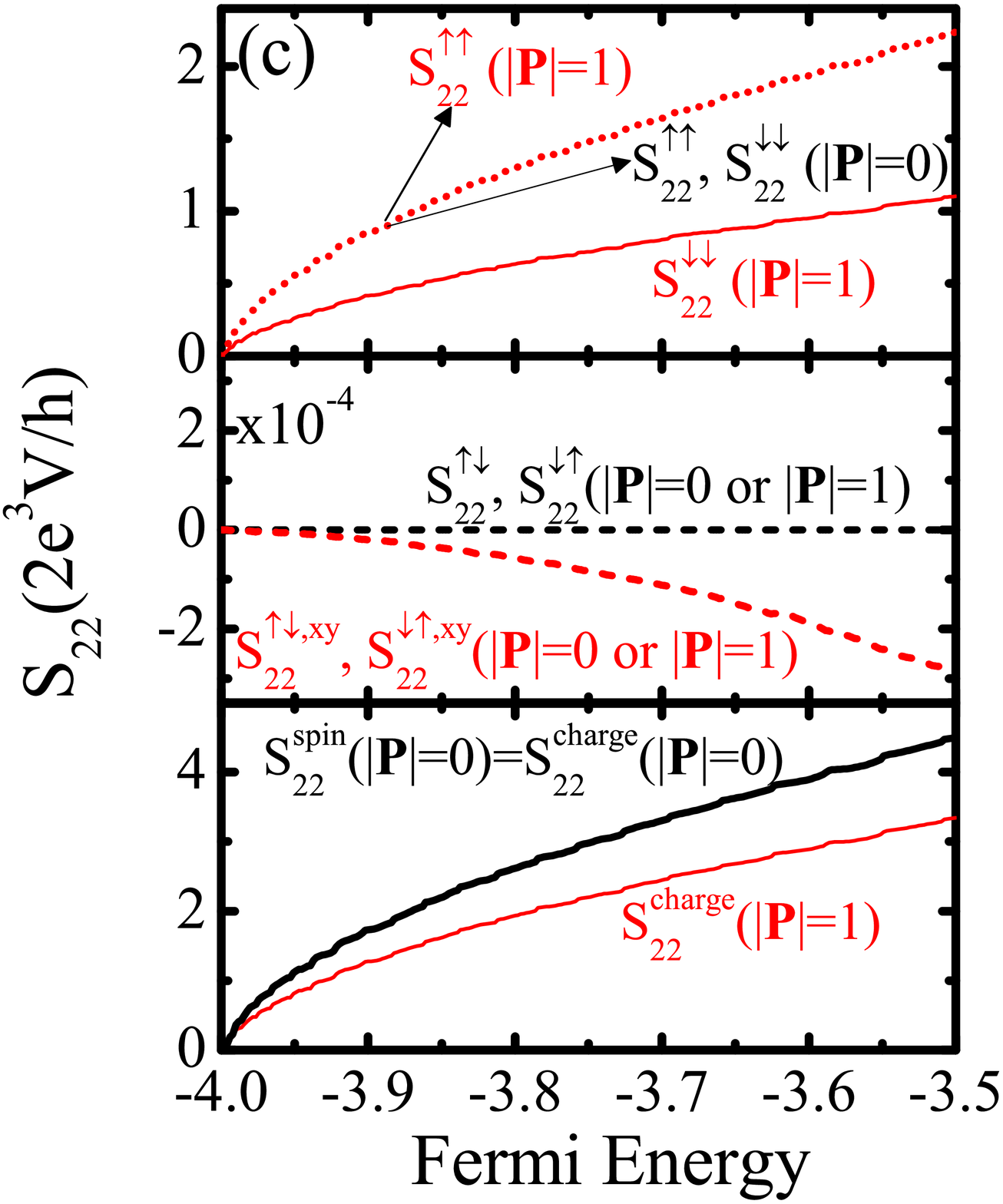,scale=0.22,angle=0}}
\caption{(Color online) Diffusive spin-resolved shot noise  in the transverse electrode 2, as well as the total shot noise of pure spin Hall current (for unpolarized $|{\bf P}|=0$ injection of $I_1$) or charge Hall current [for spin-polarized ${\bf P}=(0,0,1)$ injection of $I_1$], in {\em disordered} four-terminal 2DEGs with the Rashba SO coupling [panels (a) and (b)] or extrinsic SO scattering of strength $\lambda/\hbar =5.3$ \AA{}$^2$ [panel (c)]. In panel (b) the 2DEG sample is of the size $L_{\rm SO} \times L_{\rm SO}$, while in panels (a), (c) the sample size is fixed at $300 \, {\rm nm} \times 300 \, {\rm nm}$. The Fermi energy in panel (a) is set to allow for 23 open conducting channels [see Fig.~\ref{fig:ballistic}(a)].}\label{fig:diffusive}
\end{figure*}

In this Letter we investigate whether the information stored in the shot noise of transverse spin Hall current, as well as the noise of
associated transverse charge transport, can be used to separate different types of SO interactions driving SHE. We draw inspiration
for this approach from few recent intriguing theoretical findings:
(i) the intrinsic aspects of AHE have been directly related to the voltage shot noise~\cite{Timm2004} in four-terminal or current noise in two-terminal~\cite{Hatami2006} ferromagnetic devices; (ii) it has been argued~\cite{Erlingsson2005} that  measurements of charge currents and their auto- and cross-correlation shot noise on a  multiterminal bridge can be used to express its SHE conductance in terms of purely electrical quantities (independently of microscopic SO mechanisms); (iii) electrical measurement of the shot noise of spin-polarized charge current can sensitively probe spin precession and spin dephasing of transported electrons in two-terminal nanostructures~\cite{Dragomirova2007}. Our principal findings, summarized in Figs.~\ref{fig:ballistic} and ~\ref{fig:diffusive}, suggest that in multiterminal two-dimensional electron gas (2DEG), whose SHE can in general contain contributions from both extrinsic and intrinsic SO couplings~\cite{Nikoli'c2007}, the shot noise of spin and charge current in the transverse electrodes is substantially affected by changing
the intrinsic Rashba coupling. On the other hand, extrinsic SO scattering has virtually no effect on the shot noise in transverse electrodes.

{\em Scattering approach to shot noise of spin and charge currents in multiterminal nanostructures}---Unlike seminal arguments~\cite{Sinova2004} for the intrinsic SHE in infinite 2DEGs, where electric-field-driven acceleration of electron momenta and associated precession of spins plays crucial role, the so-called mesoscopic SHE~\cite{Nikoli'c2005e,Sheng2006b} was introduced in ballistic finite-size systems attached to multiple current and voltage probes where electric field is absent in the SO-coupled central sample (surrounded by reflectionless leads). Its description in terms of total charge $I_\alpha=I_\alpha^\uparrow + I_\alpha^\downarrow$ and conserved total spin currents $I_\alpha^{S_z}=I_\alpha^\uparrow - I_\alpha^\downarrow$ in ideal electrodes $\alpha$ (which are related to nonequilibrium spin densities within the sample and the leads~\cite{Nikoli'c2006a}) is particularly suited for the spin-dependent shot noise analysis.

To proceed with such analysis, we define correlators between spin-resolved charge currents $I^\uparrow_\alpha$ and $I^\downarrow_\beta$ in the same $\alpha = \beta$ or different $\alpha \neq \beta$ leads
\begin{equation}\label{eq:noise_time}
S_{\alpha \beta}^{\sigma \sigma^\prime} (t-t^\prime) = \frac{1}{2} \langle \delta \hat{I}_\alpha^\sigma(t)
\delta \hat{I}_\beta^{\sigma^\prime}(t^\prime) +  \delta \hat{I}_\beta^{\sigma^\prime}(t^\prime)  \delta \hat{I}_\alpha^\sigma(t) \rangle.
\end{equation}
Here $\hat{I}_\alpha^\sigma(t)$ is the quantum-mechanical operator of the spin-resolved ($\sigma=\uparrow,\downarrow$) charge current in lead $\alpha$. The current-fluctuation operator at time $t$ in lead $\alpha$ is $\delta \hat{I}_\alpha(t) = \hat{I}_\alpha(t) - \langle \hat{I}_\alpha (t) \rangle$. We use $\langle \ldots \rangle$ to denote both quantum-mechanical and statistical averaging over the states in the macroscopic reservoirs to which a mesoscopic conductor is attached via semi-infinite interaction-free leads. The spin-resolved noise power between terminals $\alpha$ and $\beta$ is  the Fourier transform of Eq.~(\ref{eq:noise_time}), $S_{\alpha\beta}^{\sigma \sigma^\prime} (\omega) = 2 \int d(t-t^\prime)\, e^{-i\omega(t-t^\prime)} S_{\alpha\beta}^{\sigma \sigma^\prime} (t-t^\prime)$, so that for charge current one gets $S_{\alpha\beta}^{\rm charge}(\omega) = S_{\alpha\beta}^{\uparrow \uparrow}(\omega) + S_{\alpha\beta}^{\downarrow \downarrow}(\omega) + S_{\alpha\beta}^{\uparrow \downarrow}(\omega) + S_{\alpha\beta}^{\downarrow \uparrow}(\omega)$ and for spin current $S_{\alpha\beta}^{\rm spin}(\omega) = S_{\alpha\beta}^{\uparrow \uparrow}(\omega) + S_{\alpha\beta}^{\downarrow \downarrow}(\omega) - S_{\alpha\beta}^{\uparrow \downarrow}(\omega) - S_{\alpha\beta}^{\downarrow \uparrow}(\omega)$.

The  scattering theory of quantum transport yields $\hat{I}_{\alpha}^{\sigma}(t)  = \frac{e}{h} \sum_{n=1}^{M} \int \!\! \int dE\,dE' \, e^{i(E-E')t/\hbar} [ \hat{a}_{\alpha n}^{\sigma \dagger}(E)\hat{a}_{\alpha n}^{\sigma}(E') \nonumber -\hat{b}_{\alpha n}^{\sigma \dagger}(E) \hat{b}_{\alpha n}^{\sigma}(E')]$ for the operator of spin-resolved charge current carried by spin-$\sigma$ electrons in terminal $\alpha$. The operator $\hat{a}^{\sigma \dagger}_{\alpha n}(E)$ [$\hat{a}^{\sigma}_{\alpha n}(E)$] creates [annihilates] incoming electrons in lead $\alpha$ which have energy $E$, spin-$\sigma$, and orbital part of their wave function is the transverse propagating mode $|n\rangle$. Similarly, $\hat{b}^{\sigma \dagger}_{\alpha n}$, $\hat{b}_{\alpha n}^\sigma$ denote spin-$\sigma$ electrons in the outgoing states. Inserted $\hat{I}_{\alpha}^{\sigma}(t)$ into Eq.~(\ref{eq:noise_time}) and Fourier transforming it leads to the following formula for the
spin-resolved  noise power spectrum
\begin{eqnarray}\label{eq:noise_power}
S_{\alpha\beta}^{\sigma\sigma'} (\omega)  & = & \frac{e^2}{h} \int dE \, \sum_{\gamma,\gamma'} \sum_{\rho,\rho'=\uparrow,\downarrow} {\rm Tr} \,  \left [{\bf A}_{\gamma\gamma'}^{\rho\rho'}(\alpha,\sigma,E,E+\hbar\omega) \right. \nonumber \\
&& \times \left. {\bf A}_{\gamma'\gamma}^{\rho'\rho}(\beta,\sigma',E+\hbar\omega,E) \right] \{f_{\gamma}^{\rho}(E)[1-f_{\gamma'}^{\rho'}(E+\hbar\omega)] \nonumber \\
&& + f_{\gamma'}^{\rho'}(E+\hbar\omega)[1-f_{\gamma}^{\rho}(E)]\}.
\end{eqnarray}
Here $f_\gamma^{\rho}(E)$ is the Fermi function of spin-$\rho$ electrons ($\rho=\uparrow,\downarrow$), kept at temperature $T_\gamma$ and spin-dependent chemical potential $\mu_\gamma^{\rho}$ in lead $\gamma$. The B\" uttiker's current matrix~\cite{Blanter2000} ${\bf A}_{\beta\gamma}^{\rho\rho'}(\alpha,\sigma,E,E')$, whose elements are $
[{\bf A}_{\beta\gamma}^{\rho\rho'}(\alpha,\sigma,E,E')]_{mn}  =  \delta_{m n} \delta_{\beta \alpha} \delta_{\gamma \alpha} \delta^{\sigma \rho} \delta^{\sigma
  \rho'}  -\sum_{k} [{\bf s}_{\alpha \beta}^{\sigma \rho\dagger}(E)]_{mk} [{\bf s}_{\alpha
  \gamma}^{\sigma \rho'}(E')]_{kn}$,
is now generalized to include explicitly spin degrees of freedom through the spin-resolved scattering matrix connecting operators $\hat{a}^{\sigma}_{\alpha n}(E)$ and $\hat{b}^{\sigma}_{\alpha n}(E)$ via $\hat{b}_{\alpha n}^{\sigma}(E)=\sum_{\beta m} [{\bf s}_{\alpha \beta}^{\sigma \sigma'}]_{nm}(E) \hat{a}_{\beta m}^{\sigma'}(E)$. In the zero-temperature limit the thermal (Johnson-Nyquist) contribution
to the noise vanishes and the Fermi function becomes a step function $f_\alpha^{\rho}(E)=\theta(E-\mu_\alpha^{\rho})$.

Evaluation of Eq.~(\ref{eq:noise_power}) for  zero-temperature and zero-frequency $S_{\alpha\beta}^{\sigma \sigma^\prime} \equiv S_{\alpha\beta}^{\sigma \sigma^\prime}(\omega=0,T=0)$ in the top lead $\alpha=2=\beta$ of a four-terminal bridge typically employed in
the analysis of the mesoscopic SHE  yields explicit  expressions for $S_{22}^{\sigma \sigma^\prime}$, $S_{22}^{\rm spin}$, and $S_{22}^{\rm charge}$ noise power [for labeling of the four leads see inset in the middle panel of Fig~\ref{fig:ballistic}(a)]. They are too lengthy to be written down here due to many terms arising from the effect of other leads on the noise in the selected lead 2. In addition, one can also compute these noise correlators for set-ups where spin-polarized charge current is injected through lead 1, thereby driving the transverse charge Hall current~\cite{Bulgakov1999} through leads 2 and 3. In this case, the magnitude $|{\bf P}|$ of the spin-polarization vector enters into Eq.~(\ref{eq:noise_power}) via the spin-dependent electrochemical potentials in the injecting lead 1,  $\mu_1^\uparrow=E_F+eV$ and $\mu_1^\downarrow=E_F+eV(1-|{\bf P}|)/(1+|{\bf P}|)$, where $E_F$ is the Fermi energy of electrons in the macroscopic reservoirs. In the collecting leads the electrochemical potentials for both spin species are the same $\mu_i^\uparrow = \mu_i^\downarrow$ ($i=2,3,4$). Such set-up is also closely related to the inverse SHE in which case $\mu_1^\uparrow=E_F+eV=\mu_2^\downarrow$ and $\mu_1^\downarrow=E_F=\mu_2^\uparrow$ (for $|{\bf P}|$=1) generates situation with longitudinal pure spin current (as injection of two counter propagating fully spin-polarized charge currents of opposite ${\bf P}$) and no net longitudinal charge current.

In general, the central 2DEG sample employed in SHE experiments~\cite{Sih2005a} can be modeled by the effective mass Hamiltonian which takes into account confining $V_{\rm conf}(y)$ and impurity $V_{\rm dis}(x,y)$ potentials, as well as intrinsic and extrinsic SO coupling effects
\begin{eqnarray}\label{eq:hamil}
\hat{H}  & = &  \frac{\hat{p}_x^2+\hat{p}_y^2}{2m^*} + V_{\rm conf}(y) + V_{\rm dis}(x,y) \nonumber \\
 && + \frac{\alpha}{\hbar}
\left( \hat{p}_y \hat{\sigma}_x  - \hat{p}_x  \hat{\sigma}_y  \right) + \lambda \left(\hat{\bm{\sigma}} \times \hat{\bf p} \right) \cdot \nabla V_{\rm dis}(x,y).
\end{eqnarray}
Here the fourth term is the intrinsic Rashba SO coupling~\cite{Winkler2003} due to structural inversion asymmetry of the quantum well [$(\hat{\sigma}_x,\hat{\sigma}_y,\hat{\sigma}_z)$ denotes the vector of the Pauli matrices, and $\hat{\bf p}=(\hat{p}_x,\hat{p}_y)$ is the momentum operator in 2D space], which is responsible for  $\Delta_{\rm SO} = 2 \alpha k_F$ spin splitting at the Fermi level ($\hbar k_F$ is the Fermi momentum). The fifth (Thomas) term is a relativistic correction to the Pauli equation for spin-$\frac{1}{2}$ particle where minuscule value of $\lambda$ in vacuum can be renormalized enormously by the band structure effects due to strong crystal potential (leading to, e.g.,  $\lambda/\hbar =5.3$ \AA{}$^2$ for GaAs~\cite{Winkler2003}). Using the unitarity of the scattering matrix, $S_{22}^{\sigma \sigma^\prime}$ as the basic noise quantity  can be expressed solely in terms of the transmission matrix ${\bf t}_{\alpha\beta}^{\sigma \sigma'}$, which is a block of the full scattering matrix determining the probability $|[{\bf t}_{\alpha\beta}^{\sigma \sigma^\prime}]_{nm}|^2$ for spin-$\sigma^\prime$  electron incident in lead $\beta$ in the orbital conducting channel $|m\rangle$ to be transmitted to lead $\alpha$ as spin-$\sigma$ electron in channel $|n\rangle$.

The  computation
of ${\bf t}_{\alpha\beta}^{\sigma \sigma'}$ for multiterminal structures described by Eq.~(\ref{eq:hamil}) is discussed in detail in Refs.~\cite{Nikoli'c2005e} through numerically exact real$\otimes$spin space Green functions for the lattice version~\cite{Nikoli'c2007} of Eq.~(\ref{eq:hamil}) which make possible to treat both weekly ($L \ll L_{\rm SO}$) and strongly ($L \ge L_{\rm SO}$) SO-coupled multichannel nanostructures, as well as its arbitrary shape and lead arrangement. Here the spin precession length $L_{\rm SO}$ (typically $L_{\rm SO} \sim 100$ nm), on which spin precesses by an angle $\pi$, plays a crucial role in the mesoscopic SHE. It is related to the Rashba coupling strength $\alpha$ through $L_{\rm SO}= \pi \hbar^2/2m^* \alpha$, and can be extracted from measurement of spin dephasing in both ballistic and diffusive systems. The spin quantization axis for $\uparrow$ and $\downarrow$ spin states is assumed to be the $z$-axis, so that all spin currents and noises in lead 2 and 3 describe the SHE response of 2DEG.


{\em Multiterminal spin and charge shot noise in ballistic 2DEG nanostructures}---In this Section and related Fig.~\ref{fig:ballistic} we assume ballistic transport [$V_{\rm dis}(x,y)=0$] through 2DEG with non-zero  $L_{\rm SO}$ due to the Rashba coupling. We recall that in two-terminal ballistic structure the stream of electrons (injected from noiseless electrodes) is completely correlated by the Pauli principle in the absence of impurity backscattering, so that corresponding shot noise vanishes $S=0$ (except at the subband edges where new conducting channels open up)~\cite{Blanter2000}. The non-zero noise $S=2Fe\langle I \rangle$  requires stochasticity of quantum-mechanical scattering off impurities or walls of chaotic cavities which, together with Pauli blocking or Coulomb interaction induced correlations, set the Fano factor $F$ (such as the well-known $F=1/3$ in diffusive conductors or $F=1/4$ in chaotic ballistic cavities for noninteracting electrons~\cite{Blanter2000}). However, in four-terminal structures in Fig.~\ref{fig:ballistic} transmission is not perfect because of the transverse leads effects (even if they do not draw current~\cite{Sukhorukov1999}), so that non-zero noise appears in the absence of SO coupling. While large Rashba coupling would introduce backscattering~\cite{Nikoli'c2005e} at the interface between the interaction-free electrodes and the sample, we find this effect not to be the crucial one for noise discussion below since similar results are obtained for the bridge where leads 1 and 4 have the same Rashba SO coupling as in the central 2DEG sample.

Since magnitude of the mesoscopic SHE increases with the sample size $L$, reaching maximum~\cite{Nikoli'c2005e,Moca2007} when $L \simeq L_{\rm SO}$, we employ the 2DEG sample of size $L_{\rm SO} \times L_{\rm SO}$ to study the dependence of the shot noise on the Fermi energy (i.e., charge density that can be changed in experiments by gate electrodes). We also assume that 2DEG is smaller than the inelastic scattering length $L_{\rm in}$ because  in larger samples electron-phonon scattering averages out the shot noise to zero~\cite{Blanter2000}. The most conspicuous feature of spin-resolved noise in Fig.~\ref{fig:ballistic} is the emergence of highly non-trivial cross-correlations between spin-resolved currents encoded by $S_{22}^{\uparrow \downarrow}=S_{22}^{\downarrow \uparrow}<0$ (more pronounced for polarized $|{\bf P}|=|(0,0,1)|=1$ injection). This stems from spin flips in the form of continuous spin precession of the $z$-axis oriented spins in the in-plane effective momentum-dependent magnetic field corresponding to the Rashba SO coupling. These cross-correlations can be manipulated by changing the Fermi energy in the case of polarized injection ($|{\bf P}|=1)$ or Rashba coupling in the case of unpolarized longitudinal current (${\bf P}=0$), thereby imprinting signature of intrinsic SO couplings on  experimentally measurable charge current noise $S_{22}^{\rm charge}$.

Another feature specific to mesoscopic manifestations of SHE, which is also exhibited by the SHE conductance $I_2^{S_z}/V$~\cite{Nikoli'c2005e,Sheng2006b}, is the appearance of sharp noise peaks in Fig.~\ref{fig:ballistic}(a) in the vicinity of
subband edges. At these energies new conducting channels in the leads become open for transport [top panel of Fig.~\ref{fig:ballistic}(a)].
Although this multiterminal noise property of ballistic conductors persists even in the absence of SO coupling, additional features of
this type can arise at energies of bound states in the cross whose mixing with propagating states via SO coupling introduces
resonances in the transmission~\cite{Bulgakov1999}.

We emphasize~\cite{Nikoli'c2005e,Nikoli'c2007} that achieving pure ($I_2=I_3=0$) spin Hall current $I_2^{S_z}$, akin to SHE in
infinite systems~\cite{Sinova2004,Guo2008}, demands to apply voltages $\mu_2=\mu_3=eV/2$ to transverse leads of the bridge biased
with $\mu_1=eV$ and $\mu_4=0$. Despite zero charge current $I_2=0$ in this case, we find non-zero fluctuations around zero average value
as encoded by $S_{22}^{\rm charge}(|{\bf P}|=1)$. This noise power increases in the same set-up, at fixed $E_F$ and with fast spin dynamics
in samples $L/L_{\rm SO} \gtrsim 2$,  by switching from unpolarized to polarized (partially $0<|{\bf P}|<1$ or fully $|{\bf P}|=1$)
injection of longitudinal current $I_1$ responsible for non-zero transverse charge Hall current~\cite{Bulgakov1999}. Note that due to
$I_2=0$ in the SHE set-up (${\bf P}=0$), we plot raw noise values (in all Figures for ease of comparison) rather than normalizing them
to $2eI_2$ (to get the Fano factors).

{\em Multiterminal spin and charge shot noise in diffusive 2DEG nanostructures}---To bring a multiterminal SHE bridge into the diffusive transport regime we introduce disorder into 2DEG and tune its strength to ensure that shot noise in lead 1 attains the Fano factor $F_{11}=S_{11}^{\rm charge}/2eI_1=1/3$. In the absence of SO coupling, noise in other three leads does not display any universal features ($F_{11}=1/3$ is expected to be independent of the impurity distribution, band structure, and shape of the conductor~\cite{Sukhorukov1999}) because of nonlocal effects---other leads contribute to the noise in the electrode $\alpha \neq 1$ making possible arbitrarily large values beyond $1/3eI_\alpha$~\cite{Sukhorukov1999}.

In the presence of disorder, one can expect both extrinsic and intrinsic contributions to $I_2^{S_z}$ whose importance (as in the case of experimental SHE systems based on 2DEG~\cite{Sih2005a}) is governed~\cite{Nikoli'c2007} by the ratio of characteristic energy scales $\Delta_{\rm SO} \tau/\hbar$ ($\hbar/\tau$ is disorder induced broadening of energy levels due to transport scattering time $\tau$). For simplicity, we analyze separately 2DEGs with dominant intrinsic [$\alpha \neq 0$, $\lambda=0$ in Fig.~\ref{fig:diffusive}(a),(b)] and extrinsic [$\alpha =0$, $\lambda/\hbar =5.3$ \AA{}$^2$ in Fig.~\ref{fig:diffusive}(c)] regimes of SHE. The most important insight brought about by Fig.~\ref{fig:diffusive} is substantial difference between the shot noise in the intrinsic and extrinsic regime, where the former exhibits non-zero cross correlations $S_{22}^{\uparrow \downarrow}=S_{22}^{\downarrow \uparrow}<0$ (as in the ballistic case, but smaller). The latter has no correlations of this
type for the $z$-axis spins and orders of magnitude smaller cross-correlation noise for the $x$- or $y$-spins whose spin currents due to $\lambda \neq 0$ are zero $I_2^{S_x}=I_2^{S_y} \equiv 0$. We also find that hypothetical increase of $\lambda$ would give orders of magnitude smaller
noise change (in fact, decrease) compared to significant spin  $S_{22}^{\rm spin}(|{\bf P}|=0)$ or charge $S_{22}^{\rm charge}(|{\bf P}|=1)$ shot noise enhancement with increasing intrinsic $\alpha$.

{\em Conclusions}---We predict that in low-dimensional mesoscopic SHE systems any intrinsic SO mechanisms involving precessing spins would lead to
significant enhancement of the shot noise of spin and charge transport, as well as to non-trivial correlations between spin-resolved
currents of opposite spin states. In contrast, the extrinsic SO scattering off impurities in 2D has virtually no effect on the zero
temperature noise. Therefore, experiment observing shot noise enhancement in the transverse electrodes upon changing the voltage
of a gate covering 2DEG would unambiguously confirm the dominance of the intrinsic contribution to the spin Hall or the charge Hall effect
(and related inverse SHE)  in multiterminal nanostructures.

\begin{acknowledgments}
This research was supported by DOE Grant No. DE-FG02-07ER46374 and NSF Grant No. ECCS 0725566.
\end{acknowledgments}

%********************references************************************************************************

%*****************************************************************


\begin{thebibliography}{10}

\bibitem{Kato2004b} Y.~K. Kato {\it et al.}, Science {\bf 306},
   1910  (2004).

\bibitem{Awschalom2007}
D.~D. Awschalom and M.~E. Flatt\' e, Nat. Phys. {\bf 3},  153  (2007).

\bibitem{Valenzuela2006} S.~O. Valenzuela and M. Tinkham, Nature {\bf 442},  176  (2006).

\bibitem{Hirsch1999}
J.~E. Hirsch, Phys. Rev. Lett. {\bf 83},  1834  (1999).


\bibitem{Sinova2004}
J. Sinova {\it et~al.}, Phys. Rev. Lett. {\bf 92},  126603  (2004).

\bibitem{Guo2008} G.~Y. Guo {\it et al.}, Phys. Rev. Lett.
  {\bf 100},  096401  (2008).

\bibitem{Nikoli'c2006a} B.~K. {Nikoli{\'c}}, L.~P. {Z{\^a}rbo}, and S. {Souma}, Phys. Rev. B {\bf
  73},  075303  (2006).

\bibitem{Zyzin2008} V. A. Zyuzin, P. G. Silvestrov, and E. G. Mishchenko, Phys. Rev. Lett. {\bf 99}, 106601 (2007).

\bibitem{Nikoli'c2005e}
B.~K. {Nikoli\'{c}}, L.~P. {Z\^{a}rbo}, and S. Souma, Phys. Rev. B {\bf 72},
  075361  (2005).

\bibitem{Sinitsyn2008}
N.~A. Sinitsyn, J. Phys.: Condens. Matter {\bf 20},  023201 (2008).

\bibitem{Kotzler2005}
J. {K\"{o}tzler} and W. Gil, Phys. Rev. B {\bf 72},  060412(R)  (2005).

\bibitem{Blanter2000}
Y.~M. Blanter and M. B{\"u}ttiker, Phys. Rep. {\bf 336},  1  (2000).

\bibitem{Mishchenko2003} E.~G. Mishchenko, Phys. Rev. B {\bf 68},  100409  (2003); A. {Lamacraft}, Phys. Rev. B {\bf 69},  081301  (2004);
W. {Belzig} and M. {Zareyan}, Phys.  Rev. B {\bf 69},  140407  (2004); K.~E. {Nagaev} and L.~I. {Glazman}, Phys. Rev. B {\bf 73},  054423  (2006).

\bibitem{Dragomirova2007} R.~L. Dragomirova and B.~K. Nikoli\'{c}, Phys. Rev. B {\bf 75},  085328
  (2007).

\bibitem{Guerrero2006} R. Guerrero {\it et~al.}, Phys. Rev. Lett. {\bf 97},  266602  (2006); S. Garzon, Y. Chen, and R. Webb, Physica E {\bf 40},  133  (2007).

\bibitem{Timm2004} C. Timm, F. von Oppen, and F. H\"{o}fling, Phys. Rev. B  {\bf 69},  115202  (2004).

\bibitem{Hatami2006} M. Hatami and M. Zareyan, Phys. Rev. B {\bf 73},  172409  (2006).

\bibitem{Erlingsson2005} S.~I. Erlingsson and D. Loss, Phys. Rev. B {\bf 72},  121310  (2005).

\bibitem{Nikoli'c2007} B.~K. Nikoli\'{c} and L.~P. Z\^{a}rbo, Europhys. Lett. {\bf 77},  47004  (2007).

\bibitem{Sheng2006b} L. Sheng and C.~S. Ting, Int. J. Mod. Phys. B {\bf 20},  2339  (2006).

\bibitem{Bulgakov1999} E.~N. Bulgakov {\it et~al.}, Phys. Rev. Lett. {\bf 83},  376  (1999).

\bibitem{Sih2005a} V. Sih {\it et~al.}, Nature Phys. {\bf 1}, 31 (2005).

\bibitem{Winkler2003} R. Winkler, {\em Spin-Orbit Coupling Effects in Two-Dimensional Electron and
  Hole Systems} (Springer, Berlin, 2003).

\bibitem{Sukhorukov1999} E. V. Sukhorukov and D. Loss, Phys. Rev. B {\bf 59}, 13054 (1999).

\bibitem{Moca2007} C.~P. Moca and D.~C. Marinescu, Phys. Rev. {\bf B 75},  035325  (2007).

\end{thebibliography}
\end{document}